\documentclass[aps,10pt,prl,twocolumn,preprintnumbers,amsmath,amssymb,nofootinbib,superscriptaddress,a4paper,showpacs,tightenlines]{revtex4-1}
\pdfoutput=1
\newcommand{\eq}[1]{\vspace{-5.5pt}\begin{equation}\hspace{2pt}#1\hspace{-0pt}\vspace{-3.5pt}\end{equation}}
\usepackage{xcolor,titlesec,tikz-cd,datetime2,graphicx,amsfonts,amsthm,bm,bbm,mathrsfs,amsmath,float,pgf,tikz,mathrsfs}
\titleformat{\section}{\centering\normalsize\normalfont\bf}{\thesection}{0em}{}
\usetikzlibrary{cd}
\definecolor{lapis}{rgb}{0.0.0470,0.2941,0.5568}
\definecolor{burgundy}{rgb}{0.5, 0.0, 0.13}
\usepackage[T1]{fontenc}
\usepackage[english]{babel}
\usepackage[utf8]{inputenc}
\usetikzlibrary{arrows}

\usepackage[colorlinks=true, linkcolor=burgundy, urlcolor=lapis, citecolor=lapis, anchorcolor=green]{hyperref}  
\newcommand{\lab}[1]{\bigg\langle{#1\bigg\rangle}}
\begin{document}
\title{{Ambitwistor Integrands from Tensionless Chiral Superstring Integrands}}
\author{\vspace{-0.25cm}\textsc{Nikhil Kalyanapuram}}
\email{nkalyanapuram@psu.edu}
\affiliation{Department of Physics and Institute for Gravitation and the Cosmos, The Pennsylvania State University, University Park PA 16802, USA}


\begin{abstract}
It is established that in the tensionless limit the chiral superstring integrand is reduced to the chiral integrand of the ambitwistor string.
\end{abstract}


\maketitle


\vspace{-15pt}
\section{Introduction}
\vspace{-14pt}
Since the twistor string formulation of scattering amplitudes in $\mathcal{N}=4$ super Yang-Mills \cite{Witten:2003nn, Roiban:2004vt, Roiban:2004yf}, much effort has been expended in developing worldsheet approaches to scattering amplitudes in more general quantum field theories. An important advance was made by the scattering equation formalism of Cachazo, He and Yuan \cite{Cachazo:2013gna,Cachazo:2013hca,Cachazo:2013iea,Cachazo:2014nsa,Cachazo:2014xea}, which realised scattering amplitudes in a large class of theories at tree level as integrals over the moduli space $\mathcal{M}_{0,n}$ of marked Riemann spheres of the form

\eq{\int_{\mathcal{M}_{0,n}}\frac{d^{n}\sigma}{\mathrm{SL}(2,\mathbb{C})}\mathcal{I}_{L}\mathcal{I}_{R}\prod^{n-3}_{i=1}\delta(E_{i})}

where the $E_{i}$ are conditions known as the scattering equations that serve to fully localise the integrand at discrete points on the moduli space. The half integrands $\mathcal{I}_{L,R}$ encode kinematical or colour degrees of freedom. The freedom in choosing these objects leads to integral formulae for a large class of quantum field theories. 

Soon after the development of the CHY formalism, it was observed that the half integrands $\mathcal{I}_{L,R}$ used in evaluating amplitudes in gauge theory and gravity\footnote{It is a fact that \emph{any} quantum field theory admits such a representation \cite{Baadsgaard:2015hia,Baadsgaard:2015ifa,Baadsgaard:2016fel}. However, the integrands giving gauge and gravity amplitudes at tree level are known to enjoy properties such as manifest color-kinematic duality \cite{Cachazo:2013gna,Bjerrum-Bohr:2016axv} that make them especially interesting.} can be equivalently computed as correlation functions of a two-dimensional theory known as the \emph{ambitwistor string} \cite{Mason:2013sva}. Since this model provided an explicit worldsheet interpretation for the CHY approach, the restriction to tree level amplitudes could be overcome. The genus expansion of the ambitwistor string has since been used to derive moduli space formulae at one-loop \cite{Adamo:2013tsa,Geyer:2015bja,Geyer:2015jch} and two-loop \cite{Geyer:2018xwu} orders.

Although the ambitwistor string provides a consistent scheme through which the half integrands in the CHY framework may be determined, its relationship with the more conventional Ramond-Neveu-Schwarz (RNS) string remains to be fully understood. While it may be formally obtained as a low-energy ($\alpha'\rightarrow 0$) limit of the RNS string \cite{Mason:2013sva}, a more careful analysis seems to indicate that the two are better related by the \emph{tensionless}  ($\alpha'\rightarrow \infty$) limit \cite{Ohmori:2015sha,Casali:2016atr}.  

In this letter, considering the latter viewpoint, by direct computation of the chiral integrand \cite{DHoker:1989cxq} of superstring NS states we demonstrate that it reduces to the corresponding half integrand in the ambitwistor string as the tensionless limit is approached. More concretely, we show that the superstring integrand (we denote the spin structure by $\delta$) takes the form

\eq{\label{eq:2}\mathcal{A}_{g,n}[\delta] = \exp(\mathrm{KN})\times\mathcal{I}^{\alpha'}_{g,n}[\delta]
}
(where $\mathrm{KN}$ generalises the Koba-Nielsen factor to higher loops) such that in the limit of infinite $\alpha'$, $\mathcal{I}^{\alpha'}_{g,n}[\delta]$ equals the chiral half integrand computed by the ambitwistor string.

\vspace{-15pt}
\section{Holomorphic Factorization of Superstring Scattering Amplitudes}
\vspace{-14pt}

The scattering of $n$ NS states in superstring perturbation theory is defined by a formal integral over the supermoduli space $\mathfrak{M}_{g,n}$ (with measure $d\mu_{g,n}$) of super Riemann surfaces with $n$ NS punctures

\eq{\int_{\mathfrak{M}_{g,n}}d\mu_{g,n}\langle{ |\delta(H_{A}|B)|^{2}\rangle}_{B,C}\times \mathcal{O}_{n}}
where $H_{A}$ is a basis of Beltrami superdifferentials and $B$ and $C$ are ghost superfields encoding the $bc$ and $\beta\gamma$ systems such that

\eq{B(z,\theta) = \beta(z)+ \theta b(z),}
and 

\eq{C(z,\theta) = c(z)+ \theta \gamma(z).}
We compute the quantity $\mathcal{O}_{n}$ by making use of the chiral splitting theorem due to d'Hoker and Phong \cite{DHoker:1989cxq}, which says

\eq{\begin{aligned}
    \mathcal{O}_{n} = \int_{\mathbb{R}^{10g}}d^{10}p_{I} &\bigg|\bigg\langle\exp\left(\frac{i}{\alpha'}\int\chi(z)\psi^{\mu}\partial x^{\mu}(z)\right)\times\\
    &\prod_{i}\mathcal{V}(z_{i},\theta_{i},k_{i},\epsilon_{i})\bigg\rangle\bigg |^{2}
\end{aligned}}
where $\chi(z)$ is the gravitino, parametrising odd moduli on the supermoduli space and

\eq{\begin{aligned}
     &\mathcal{V}(z_{i},\theta_{i},k_{i},\epsilon_{i}) = \int d\widetilde{\theta}_{i} \exp\bigg(ik^{\mu}_{i}x^{\mu}_{+}(z_{i})+\\
     &\frac{2i}{\alpha'}\theta_{i}\widetilde{\theta}_{i}\epsilon^{\mu}_{i}\partial x^{\mu}_{+}(z_{i}) + \theta_{i}k^{\mu}_{i}\psi^{\mu}(z_{i})+\widetilde{\theta}_{i}\epsilon^{\mu}_{i}\psi^{\mu}(z_{i})\bigg).
\end{aligned}}
The chiral fields $x_{+}(z)$ and $\psi(z)$ are purely holomorphic and obey (given a spin structure $\delta$) the operator product expansions

\eq{x^{\mu}_{+}(z)x^{\nu}_{+}(z') \sim -\eta^{\mu\nu}\alpha'\ln(E(z,z'))}
and

\eq{\psi^{\mu}(z)\psi^{\nu}(z') \sim \eta^{\mu\nu}S_{\delta}(z,z').}
where $E(z_{i},z_{j})$ is the prime form on a genus $g$ Riemann surface and $S_{\delta}(z_{i},z_{j})$ is the genus $g$ Szego kernel for spin structure $\delta$. The loop momenta $p_{I}$ are defined as monodromies of the chiral $\partial x_{+}$ fields around the $\mathfrak{A}_{I}$ cycles of the Riemann surface 

\eq{ \oint_{\mathfrak{A}_{I}}\partial x^{\mu}_{+}(z)dz = -i\alpha' p^{\mu}_{I}.}
This condition is satisfied by performing the effective replacement 

\eq{\partial x^{\mu}(z) \rightarrow \partial x^{\mu}(z) -i\alpha' p^{\mu}\omega_{I}(z).}
With these definitions, we  define the \emph{chiral correlation function}

\eq{\label{eq:11}
    \begin{aligned}
    \mathcal{A}_{g,n}[\delta] =  &  \bigg\langle\prod_{A}\delta(\langle{H_{A}|B\rangle})\exp\left(\frac{i}{\alpha'}\int\chi(z)\psi^{\mu}\partial x^{\mu}(z)\right)\\&\times\prod_{i}\mathcal{V}(z_{i},\theta_{i},k_{i},\epsilon_{i})\bigg\rangle_{B,C,x_{+},\psi}.
    \end{aligned}}
The full superstring integrand is then expressed as a sum over two complex conjugate copies thereof

\eq{\mathcal{I}_{g,n} = \sum_{\delta,\delta'}\eta_{\delta}\widetilde{\eta}_{\delta'}\mathcal{A}_{g,n}[\delta]\overline{\mathcal{A}_{g,n}[\delta]}.}
where the complex conjugate chiral integrand is be evaluated at loop momenta $-p_{I}$. The constants $\eta_{\delta}$ and $\widetilde{\eta}_{\delta'}$ take values $\pm 1$ and perform the GSO projection based on the combination chosen. 

\vspace{-15pt}\section{Finding the Chiral Correlator on the Supermoduli Space}\vspace{-14pt}
We start with the evaluation of (\ref{eq:11}) on $\mathfrak{M}_{g,n}$. In doing this, we note that we will implicitly absorb the integrals over the auxiliary Grassmann parameters $\widetilde{\theta}_{i}$ into the measure of $\mathfrak{M}_{g,n}$. This is only done to make the resulting expressions more compact. 

We begin by computing the matter part of the chiral correlator, defined intrinsically on the supermoduli space $\mathfrak{M}_{g,n}$. To do this we integrate over the chiral $x_{+}(z)$ and $\psi(z)$ in turn. The $x_{+}(z)$ integration introduces additional terms quadratic in $\psi(z)$. The $\psi(z)$ integral is then performed using the effective propagator

\eq{\begin{aligned}
     \hat{S}_{\delta}(z,z') = &S_{\delta}(z,z') +\\
     &\frac{1}{\alpha'}\int S_{\delta}(z,w)K(w,v)S_{\delta}(v,z')dwdv
\end{aligned}}
where

\eq{K(w,v) = \chi(w)\partial_{w}\partial_{v}\ln(E(w,v))\chi(v).}
This modified contraction is effected by factoring out 

\eq{\mathcal{X}_{\mathrm{PCO}} =  \lab{\exp\left(\frac{i}{\alpha'}\int\chi(z)\partial x_{+}(z)\cdot\psi(z)dz\right)}.}
When this is done, the chiral correlator is evaluated as a conventional Gaussian integral. The result is given by 

\eq{\mathcal{X}_{\mathrm{PCO}} \times \exp\left(\mathrm{KN} + H_{0,0} + H_{0,1}+ \mathcal{O}\left(\frac{1}{\alpha'}\right)\right)}
where

\eq{\begin{aligned}
      \mathrm{KN} = &\alpha'\sum_{i,I}k_{i}\cdot p_{I}\int^{z_{i}}_{P}\omega_{I}(z)dz + \\
      &\alpha'\sum_{i\neq j}\frac{1}{2}k_{i}\cdot k_{j}\ln (E(z_{i},z_{j}))
\end{aligned}}
and

\eq{\begin{aligned}
     H_{0,0} = &\int \chi(z)\chi(z')P(z)\cdot P(z')S_{\delta}(z,z')dzdz' \\
     &+2\sum_{i}\bigg(\int [\chi(z)\theta_{i}P(z)\cdot k_{i}S_{\delta}(z,z_{i})+\\
     &\chi(z)\widetilde{\theta}_{i}P(z)\cdot\epsilon_{i}S_{\delta}(z,z_{i})]dz\bigg)+\\
     &\sum_{i\neq j}[\theta_{i}\theta_{j}k_{i}\cdot k_{j}S_{\delta}(z_{i},z_{j})
     + \widetilde{\theta_{i}}\widetilde{\theta}_{j}\epsilon_{i}\cdot\epsilon_{j}S_{\delta}(z_{i},z_{j})\\
     &-2\theta_{i}\widetilde{\theta}_{j}\epsilon_{j}\cdot k_{i}S_{\delta}(z_{j},z_{i})] + \sum_{i}2\theta_{i}\widetilde{\theta}_{i}\epsilon_{i}\cdot P(z_{i}). 
\end{aligned}}
The quantity $H_{0,1}$ vanishes for even spin structure. For odd spin structure it is given by

\eq{\begin{aligned}
     H_{0,1} = &\int\chi(z)\psi^{(0)}\cdot P(z)dz +\\ &\sum_{i}[\theta_{i}k_{i}\cdot\psi^{(0)}+
     \widetilde{\theta}_{i}\epsilon_{i}\cdot\psi^{(0)}]
\end{aligned}}
where $\psi^{0}$ is the (ten dimensional) zero mode of the worldsheet fermion. In these expressions

\eq{ P^{\mu}(z) = \sum_{i}k^{\mu}_{i}\ln(E(z,z_{i})) + \sum_{I}p^{\mu}_{I}\omega_{I}(z). }
Note that $\omega_{I}$ are Abelian differentials of the first kind.

We now make contact with the representation (\ref{eq:2}) by pointing out that the factor $\mathrm{KN}$ is not dependent on any Grassmann valued variables. Accordingly, we factor it out, leaving behind $H_{0,0}$, $H_{0,1}$ and terms subleading in $\frac{1}{\alpha'}$ contributing to $\mathcal{I}^{\alpha'}_{g,n}[\delta]$ 

\vspace{-15pt}
\section{Comparison to the Ambitwistor String}
\vspace{-14pt}
To compare our result to the corresponding chiral integrand in the ambitwistor string, we need to reduce our result to an expression on the ordinary moduli space $\mathcal{M}_{g.n}$. However, projecting down to the ordinary moduli space from the supermoduli space isn't always possible\footnote{It is known that $\mathfrak{M}_{g}$ is not projected \cite{Donagi:2013dua,Donagi:2014hza} while the question remains unanswered in the presence of punctures.}. Due to the general difficulties involved in projecting down to $\mathcal{M}_{g,n}$, it is best to work in small enough neighbourhoods $\mathcal{U}\in \mathcal{M}_{g,n}$ onto which holomorphic projections always exist\footnote{I am grateful to Seyed Faroogh Moosavian and Edward Witten for discussions on this point.}.

Let us then take some (arbitrary) point $\Sigma\in \mathcal{M}_{g,n}$. For a small enough neighbourhood $\mathcal{U}_{\Sigma}$, the moduli of $\mathfrak{M}_{g,n}$ can be parametrised by the ordinary bosonic moduli and a gravitino $\chi(z)$ that can be expanded as \cite{DHoker:2002hof}

\eq{\label{eq:19}
    \chi(z) = \sum_{i}^{N_{odd}}\chi_{\alpha_{i}}\delta(z-z_{\alpha_{i}})}
where the insertion points $z_{\alpha_{i}}$ are holomorphic functions of the bosonic moduli\footnote{The requirement that the PCOs depend on the moduli is due to the presence of spurious singularities. See \cite{Witten:2012bh,Sen:2014pia,Moosavian:2017fta} for discussions of this issue.}. The number $N_{\mathrm{odd}}$ of Grassmann directions spanned by the $\chi_{\alpha_i}$ is $0$ for genus zero and genus one with even spin structure. For genus one with odd spin structure $N_{\mathrm{odd}} = 1$ and for genus $g\geq 2$ we have $N_{\mathrm{odd}} = 2g-2$. 

In this local patch, the basis for the Beltrami superdifferentials is 

\eq{H_{A} = (\mu_{a}|\delta_{z,z_{\alpha_i}}),}
where the $\mu_{a}$ are the $N_{\mathrm{even}}$\footnote{For genus $g=0$ $N_{\mathrm{even}}=0$, for $g=1$ $N_{\mathrm{even}}=1$ and for $g\geq 2$ $N_{\mathrm{even}}=3g-3$.} ordinary Beltrami differentials labelling deformations of the bosonic moduli and $\delta_{z,z_{\alpha_i}}$ are $N_{\mathrm{odd}}$ delta functions evaluating at $z_{\alpha_i}$. 

We have shown that in this set up, upon integrating away the fermionic degrees of freedom, the matter part of the chiral correlation function and the ghost partition function precisely match the corresponding results of the ambitwistor string \cite{Geyer:2018xwu}. Putting these together, we have proved that the chiral correlator of the RNS superstring reduces to the corresponding chiral half integrand of the ambitwistor string in the $\alpha'\rightarrow \infty$ limit. Concretely we have (specialising to the case of even spin structure)

\eq{\label{eq:23}
    \mathcal{I}^{\alpha'}_{g,n}[\delta] = \mathcal{Z}_{gh}[\delta]\left(\mathrm{Pf}\Psi_{g,n}[\delta] + \mathcal{O}\left(\frac{1}{\alpha'}\right)\right).}
Here, $\mathcal{Z}_{gh}[\delta]$ is the ghost partition function

\eq{
\begin{aligned}
\mathcal{Z}_{gh}[\delta] = \frac{1}{Z^{5}}\bigg\langle&\prod^{N_{\mathrm{even}}}_{i=1}\langle{\mu_{i}|b\rangle}\prod^{N_{\mathrm{odd}}}_{i=1}\delta(\beta(z_{\alpha_i}))\\
&\prod^{N_B}_{i=1}c(z_{a_i})\prod^{N_F}_{i=1}\delta(\gamma(z_{b_{i}}))\bigg\rangle_{\delta}
\end{aligned}}
where $\langle{\mu_{i}|b\rangle}$ is used to denote the inner product between the Beltrami differential $\mu_{i}$ and the field $b$ and it is to be understood that the expectation value is to be taken with respect to the $bc$ and $\beta\gamma$ systems. $Z$ is the chiral scalar partition function (see for example \cite{Geyer:2018xwu,DHoker:2001jaf} for an explicit representation in terms of the prime form and theta functions on $\mathcal{M}_{g,n}$). $N_{B}$ and $N_{F}$ are only nonzero at genus zero and one\footnote{At $g=0$, $N_B = 3$ and $N_{F}=2$, at $g=1$ for even spin structure $N_B = 1$ and $N_F = 0$ and for odd spin structure $N_B = N_F = 1$.}, corresponding to insertions of $c$ and $\gamma$ fields to account for superconformal Killing vectors. The $z_{a_i}$ and $z_{b_i}$ are arbitrarily chosen external punctures, not to be confused with PCO insertions.  $\mathcal{Z}_{gh}[\delta]$ is computed by bosonisation \cite{Verlinde:1986kw,Verlinde:1987sd}. 

The quantity $\Psi_{g,n}$ is a matrix, given by\footnote{In evaluating the Pfaffian in (\ref{eq:23}), it is defined after removing a pair of rows and columns and one row and column corresponding to external states respectively at genus zero and genus one with odd spin structure. This is due to the fact that $N_{F}$ states with marked points $z_{b_{i}}$ are in the $-1$ picture.}

\eq{\Psi_{g,n} = \begin{pmatrix}
    \mathbf{A}&-\mathbf{C}^{T}\\
    \mathbf{C}&\mathbf{B}\end{pmatrix}}
where

\eq{\begin{aligned}
    &\mathbf{A}_{\alpha_i\alpha_j} = P(z_{\alpha_i})\cdot P(z_{\alpha_{j}})S_{\delta}(z_{\alpha_{i}},z_{\alpha_{j}}),\\
    &\mathbf{A}_{\alpha_i i} = P(z_{\alpha_i})\cdot k_{i}S_{\delta}(z_{\alpha_i},z_{i}), \\
    &\mathbf{C}_{\alpha_i j} = P(z_{\alpha_i})\cdot \epsilon_{i}S_{\delta}(z_{i},z_{\alpha_i}),\\
    & \mathbf{A}_{ij} = k_{i}\cdot k_{i}S_{\delta}(z_{i},z_{j}),\;\; \mathbf{B}_{ij} = \epsilon_{i}\cdot\epsilon_{j}S_{\delta}(z_{i},z_{j}),\\
    &\mathbf{C}_{ij} = \epsilon_{i}\cdot k_{j}S_{\delta}(z_{i},z_{j}),\;\; \mathbf{C}_{ii} = -k_{i}\cdot P(z_{i}).
\end{aligned}}
Indeed, the result (\ref{eq:23}) is in agreement with the genus $g$ chiral correlator defined in \cite{Geyer:2018xwu} after taking the $\alpha'\rightarrow 0$ limit is taken. We remark that the result is analogous for the case of odd spin structure; the chiral correlator is resolved into the ambitwistor correlator followed by corrections of order $\frac{1}{\alpha'}$ \cite{1851730}.

Regarding this choice of local projection, there is one subtlety that to our knowledge has not been explicitly considered in previous analyses of the ambitwistor string. The choice of PCO insertions must be made in a manner that avoids spurious singularities. Indeed, this can always be done locally by picking PCOs that depend holomorphically on the bosonic moduli. In piecing together such local descriptions, it is not known if a globally consistent prescription to perform localisation in the ambitwistor string can be found. In practice however, it turns out that such a selection of PCOs exists for the ambitwistor string up to two loop order \cite{Geyer:2018xwu,1851730} as a result of the fact that the integral ultimately localises on a discrete set of points in $\mathcal{M}_{2,n}$. Proving this at higher genus requires further study.

For readers familiar with the picture changing operator formalism, we point out that what we have done basically is to derive the ansatz due to Friedan, Martinec and Shenker \cite{Friedan:1985ge} using the language of supermoduli spaces. Historically, it was observed that this ansatz was poorly defined and led to amplitudes dependent on the choice of PCO insertions \cite{Verlinde:1987sd}. This is due to the fact that the projection was done globally, which as we have noted is generically not possible. It is in accordance with this that we have chosen to work in small neighbourhoods. In the context of the full superstring, local descriptions must be carefully glued together \cite{Sen:2014pia,Sen:2015hia}. Accordingly, it should be kept in mind that we have proved that the integrand $\mathcal{I}^{\alpha'}_{g,n}[\delta]$ matches the half integrand in the ambitwistor string in the tensionless limit given a specific local configuration of picture changing operators. The subtleties involved with gluing might however pose a problem if we actually want to perform the integration over the moduli space for the ambitwistor string at higher genus.

\vspace{-15pt}
\section{Conclusion}
\vspace{-14pt}
In this letter, we have proved the claim that the chiral superstring integrand in the RNS formalism is resolved into a Koba-Nielsen term and an $\alpha'$ dependent integrand that reduces to the half integrand of the ambitwistor string in the tensionless limit. We made use of the chiral splitting theorem due to d'Hoker and Phong to define a chiral correlation function for $n$ NS states, which we evaluated on the supermoduli space $\mathfrak{M}_{g,n}$. Due to potential issues involving non-projectedness at higher genus, we had to work on small neighbourhoods in $\mathcal{M}_{g,n}$, onto which a projection is possible. Evaluating the chiral correlation function on the ordinary moduli space in this fashion, we see that in the limit of vanishing tension the chiral correlator reproduced the chiral correlation function of the ambitwistor string.

A full understanding of the tensionless limit of RNS superstrings and its relation to the ambitwistor string requires clarifying the behaviour of the higher genus Koba-Nielsen factor in the Gross-Mende limit \cite{Gross:1987ar,Gross:1987kza}. Due to an infinite number of saddle points on the universal covering space of $\mathcal{M}_{g,n}$, the na\"ive sum over solutions of the scattering equations cannot be taken (see \cite{Mizera:2019vvs} for the complications arising even at genus zero). Understanding this limit on the non-separating divisors of $\mathcal{M}_{g,n}$ would be especially relevant in understanding the $\alpha'\rightarrow 0$ limit (see \cite{Bjerrum-Bohr:2014qwa} for a discussion of the genus zero case.). The interplay (and potential equivalence or inequivalence) of the two limits would have important implications for the duality between colour and kinematics \cite{Mizera:2019gea,Mizera:2019blq}.
\vspace{-15pt}
\section*{Acknowledgements}
\vspace{-15pt}
I thank Jacob Bourjaily for encouragement and constructive comments that considerably improved the draft. I have benefited from exchanges with Yvonne Geyer, Alok Laddha, Sebastian Mizera, Ricardo Monteiro, Seyed Faroogh Moosavian, Oliver Schlotterer and Edward Witten. I am especially grateful to Seyed Faroogh Moosavian for early conversations on chiral splitting which motivated this analysis. This project has been supported by an ERC Starting Grant (No. 757978) and a grant from the Villum Fonden (No. 15369).

\bibliographystyle{utphys}
\bibliography{v1.bib}

\end{document}